\documentclass[11pt,twoside]{article}
\usepackage{graphicx,epsfig,natbib,epstopdf}
\usepackage{CS18}
\begin{document}
%
%
%
\title{Empirical Limits on Radial Velocity \\ Planet Detection for Young Stars}
%
%
\author{Lynne Hillenbrand$^{1}$, 
Howard Isaacson$^{2}$, 
Geoffrey Marcy$^{2}$, 
Scott Barenfeld$^{1}$
Debra Fischer$^{3}$
Andrew Howard$^{4}$
}
\affil{$^1$ Department of Astronomy; California Institute of Technology; Pasadena, CA 91125, USA}
\affil{$^2$ Department of Astronomy; University of California; Berkeley, CA 94720, USA}
\affil{$^3$ Department of Astronomy; Yale University; New Haven, CT 06510, USA}
\affil{$^4$ Institute for Astronomy; University of Hawaii; Honolulu, HI, 96822, USA}
\begin{abstract}
%
%
We report initial results from our long term search using precision radial velocities for planetary-mass companions located within a few AU of stars younger than the Sun. Based on a sample of $>$150 stars, we define a floor in the radial velocity scatter, $\sigma_{RV}$, as a function of the chromospheric activity level $R'_{HK}$. This lower bound to the jitter, which increases with increasing stellar activity, sets the minimum planet mass that could be detected. Adopting a median activity-age relationship reveals the astrophysical limits to planet masses discernable via radial velocity monitoring, as a function of stellar age.
Considering solar-mass primaries having the mean jitter-activity level, when they are younger than 100 / 300 / 1000 Myr, the stochastic jitter component in radial velocity measurements restricts detectable companion masses to $>$ 0.3 / 0.2 / 0.1 $M_{Jupiter}$. These numbers require a large number -- several tens -- of radial velocity observations taken over a time frame longer than the orbital period. Lower companion mass limits can be achieved for stars with less than the mean jitter and/or with an increased number of observations.

\end{abstract}
%
%
%
%
%
\section{The Stellar Noise Problem}

While young, stars are rapidly rotating which causes ``activity" manifest 
as chromospheric and coronal emission.  
The fractional flux emitted in the Ca\ II $H$ and $K$ line cores 
relative to the stellar bolometric flux is known as $R'_{HK}$; see 
e.g. Noyes et al. (1984) for the formalism.
Empirical rotation-activity correlation is illustrated in e.g. White et al. (2007) and for our sample in Hillenbrand et al. (2015).  Activity increases with increasing rotation at slow rotation speeds ($<$10-15 km/s) until becoming ``saturated" at higher values of v sin{\it i}.

\begin{figure}[ht!]
\centering
\includegraphics[angle=0,width=1.02\textwidth]{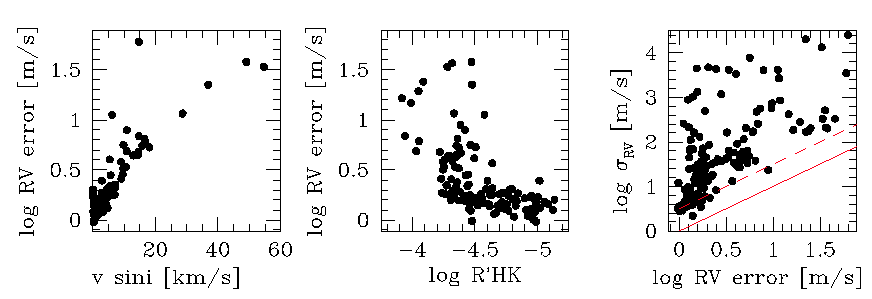}
\caption{Empirical illustration within our data set of the measured loss of radial velocity precision (increasing RV error) with increasing stellar rotation (left panel) and increasing stellar activity (middle panel).  Right panel shows the consequences on the dispersion in radial velocity measurements over time.  Red lines indicate the 1:1 and 3:1 relationships.  Points above the red lines indicate stars with significant radial velocity variations due to the combined effects of (1) stellar jitter and (2) planetary, brown dwarf, and stellar companions.
}
\end{figure}

Rapid rotation also broadens spectral absorption lines, 
making precision radial velocities harder to measure.
Separately, there is ``jitter" in observed radial velocities 
as the temperature structure of the projected stellar disk changes 
on time scales associated with photospheric granulation/convection,
rotation periods and activity cycles.
Figure 1 illustrates the relations among RV precision and 
rotation-activity-jitter
for our sample stars.

\section{Young Stars and the California Planet Search}

\begin{figure}[b!]
\centering
\vskip-0.25truein
\includegraphics[angle=0,width=0.65\textwidth]{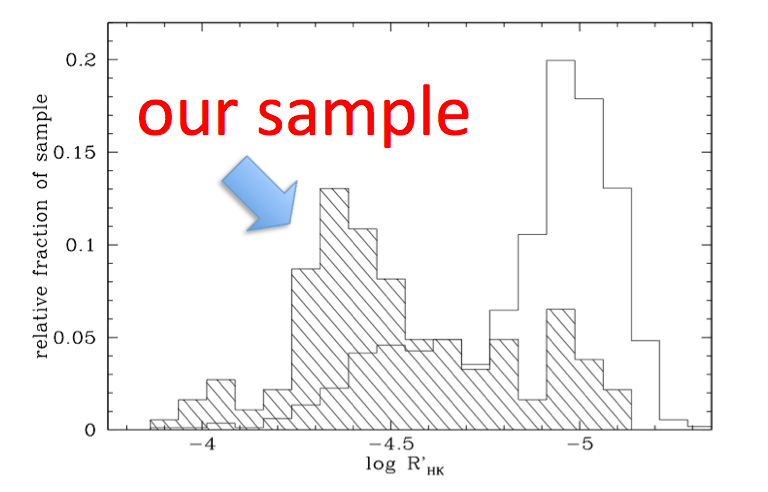}
\caption{Histogram of log $R'_{HK}$ values for 1638 FGK stars as reported in Isaacson \& Fischer (2010), representing the overall CPS sample (open histogram), compared to that of the 171 stars discussed here (shaded histogram), nearly all of which are included in the open histogram as well. Histograms are normalized by the total number of stars in each sample. Activity level increases to the left.
}
\end{figure}

Within the context of the broader California Planet Search program (e.g. Wright et al. 2012), a sample of $>$150 solar-neighborhood stars having estimated masses 0.8-1.2 $M_\odot$, and ages younger than the Sun have been monitored. The program began in 2002, with some stars added in 2004 and others in 2008. The targets were drawn from stars for which detailed information on the presence and properties of dusty disks is available. Included are stars from the FEPS (Meyer et al. 2006) and other debris disk surveys, where dust presence indicates the existence of already formed planets, and selected very young stars associated with regions of recent star formation, possibly still in the process of forming planets.

Our sample stars are young and active relative to the typically quieter stars that are the focus of much of the ongoing radial velocity planet search work;
see Figure 2.

\section{Correlation of Radial Velocity Jitter and Stellar Activity}

As illustrated in Figure 3, there is a direct correlation between the rms scatter in the observed radial velocity time series and the mean chromospheric activity level.  A linear fit suggests activity-induced radial velocity jitter of 195 m/s or more for very high activity levels, around log $R'_{HK} = -4$, and just 3 m/s (near the 3$\sigma$ measurement errors) for lower activity levels, around log $R'_{HK} = -5$.

\begin{figure}[hb!]
\centering
\includegraphics[angle=0,width=0.60\textwidth]{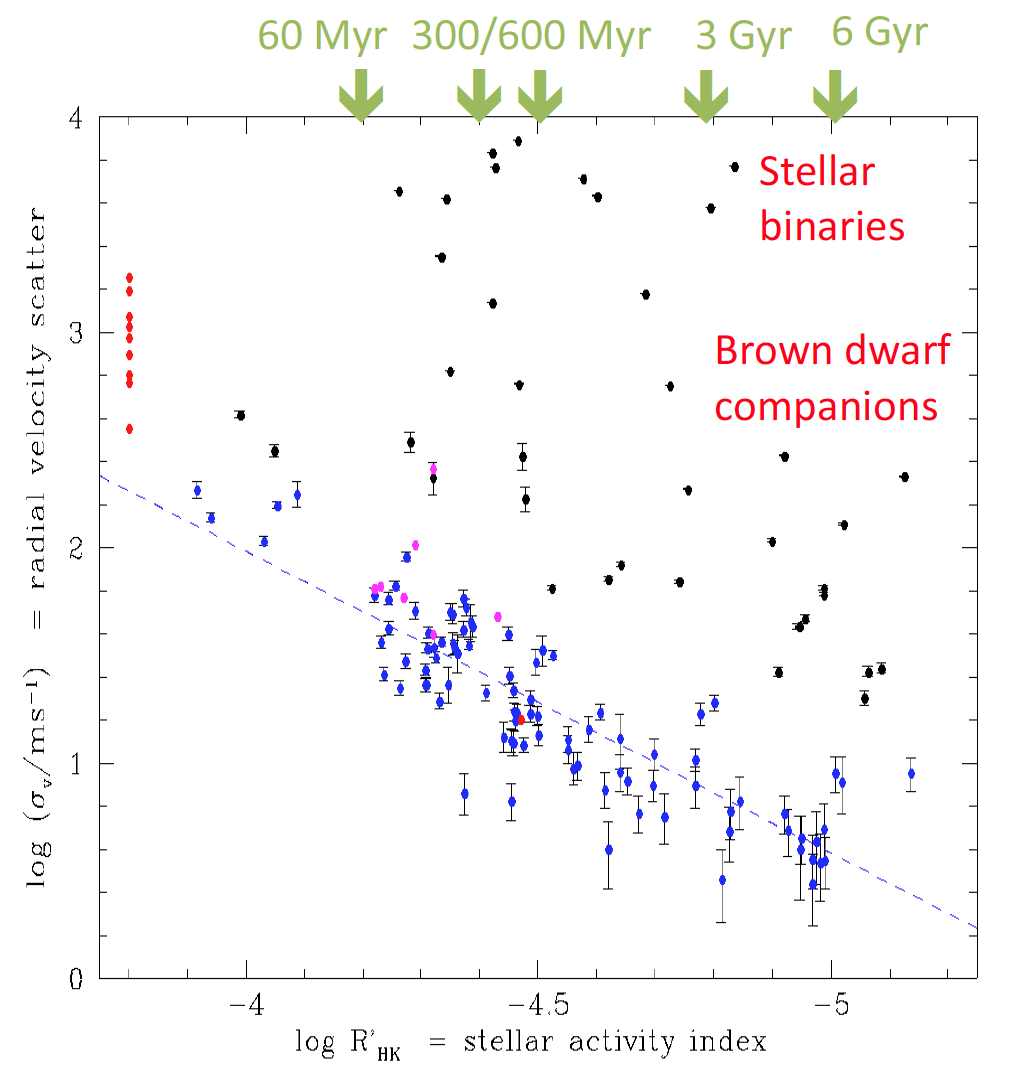}
\caption{The run of observed radial velocity rms, $\sigma_{RV}$, with stellar activity level, $R'_{HK}$. Dashed line is a linear fit to our data (blue points) after three iterations of rejecting 3$\sigma$ outliers (black points). For comparison, also shown are the rms of measured radial velocities for (red points) the 9 T Tauri stars studied by Crockett et al. (2012) at  an arbitrarily chosen activity level of log $R'_{HK} = -3.8$, and the median for Hyades stars reported by Paulson et al. (2004), adopting the median log $R'_{HK} = -4.47$ from Mamajek \& Hillenbrand (2008); at intermediate activity levels data (magenta points) from Lagrange et al. (2013) are shown.
}
\end{figure}

\section{Comparison to Previously Derived Jitter-Activity Relationships}

Our young stars are more active than those investigated in previous pursuit of jitter-activity scaling. Figure 4 illustrates how various of these relationships -- all derived using inactive star samples -- would propagate to the more active stars considered here.  Our relation is much steeper at the active end.

\begin{figure}[ht!]
\centering
\vskip-0.6truein
\includegraphics[width=0.80\textwidth]{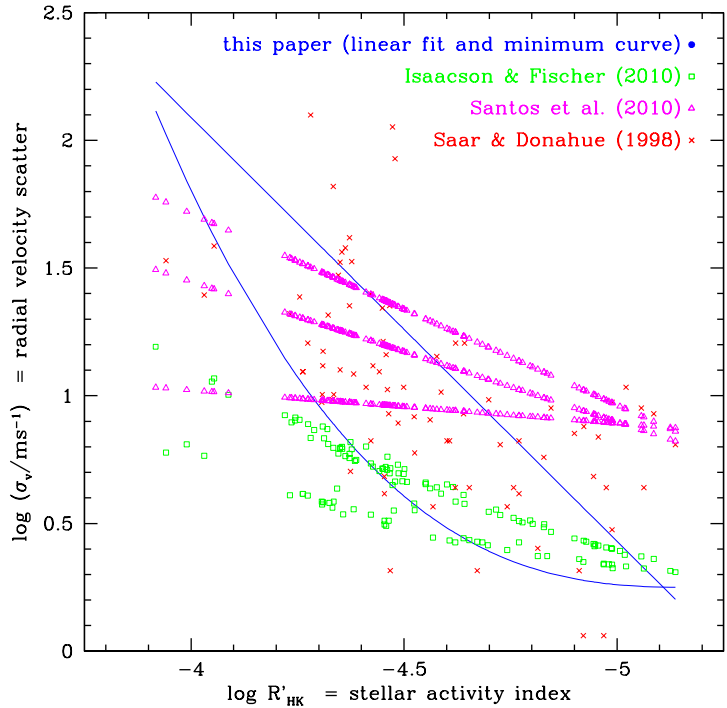}
\vskip-0.9truein
\caption{Comparison of various jitter-activity relations. Symbols are plotted as individual stars in our sample having the observed activity level but using the activity-jitter prediction of Isaacson \& Fischer (2010) in green, Santos et al. (2010) in magenta, both with three different relations for each of F,G,K stars, and the rotation-jitter prediction of Saar \& Donahue (1997) in red. None of the previous relations is in good agreement with our derived relations, shown in blue. The rotation-based scatter (red) is broader than that exhibited by our sample stars. The S-based predictions have much shallower slopes than we find, with the Isaacson \& Fischer (2010) relation underestimating the typical radial velocity jitter but matching the minimum well up to -4.4, and the Santos et al. (2010) relations a generally poor match.
}
\end{figure}

\section{Implications for RV Planet Detection}

The empirically determined radial velocity variation, $\sigma_{RV}$, as a function of activity level, $R'_{HK}$, can be turned in to a minimum detectable companion mass.  We utilize the formalism given by Narayan et al. (2005), following Cumming (2004) and Nelson \& Angel (1998).

For active stars, when the number of observations exceeds $\sim$20, meaningful limits can begin to be placed on planetary mass companions having orbits shorter than the duration of the time series. Currently we reach sub-Jupiter mass sensitivity for roughly few AU orbits, as illustrated in Figure 5.

\begin{figure}[ht!]
\centering
\includegraphics[angle=0,width=1.02\textwidth]{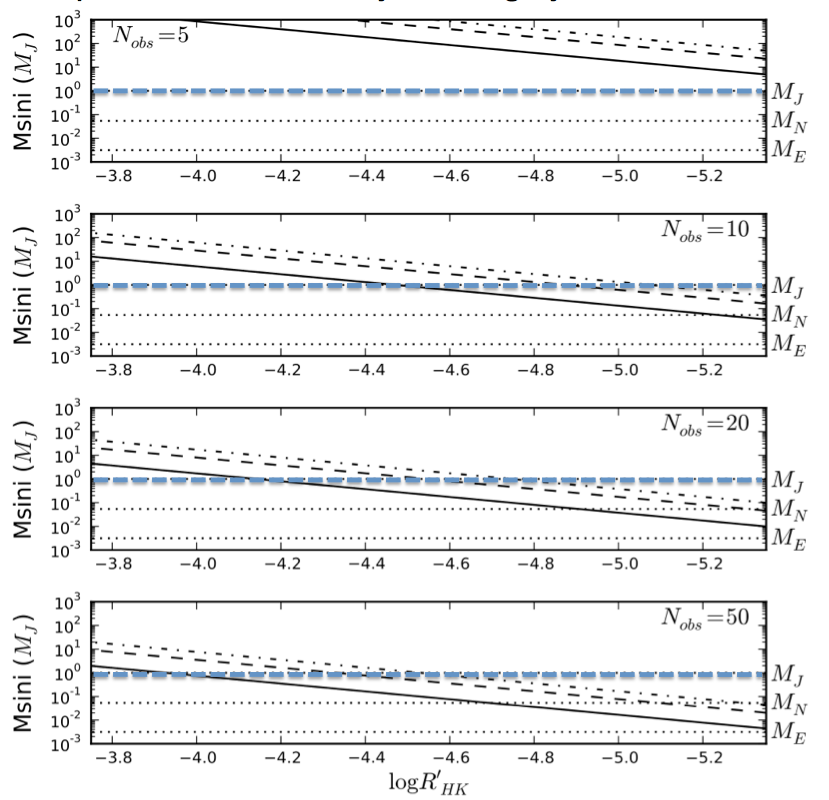}
\caption{Minimum mass of a planet with a 50\% chance of being detected in a circular orbit around a 1 $M_\odot$ star, as a function of stellar activity. Activity is translated to radial velocity noise using our empirical relationship, and then to mass based on simulations. The plot panels are for different numbers of radial velocity measurements (Nobs) and within each plot, the solid, dashed, and dotted-dashed lines correspond to planet periods of 1, 10, and 1000 days. The horizontal dotted lines show the masses of Jupiter (in blue), Neptune, and Earth.
}
\end{figure}


%
%

%
%
%

%
%

%
%




\normalsize

\newpage
\begin{references}

Crockett, C.~J., 
Mahmud, N.~I., Prato, L., et al.\ 2012, ApJ, 761, 164 

Cumming, A.\ 2004, MNRAS, 354, 1165 

Hillenbrand, L.A., Isaacson, H., Barenfeld, S., Marcy, G., Fischer, D., Howard, A. \ 2015, ApJ, submitted soon. 

Isaacson, H., \& Fischer, D.  \ 2010, ApJ, 725, 875

Lagrange, A.-M., Meunier, N., Chauvin, G., et al.\ 2013, A\&A, 559, A83 

Mamajek, E.~E., \& Hillenbrand, L.~A.\ 2008, \apj, 687, 1264 

Meyer, M.~R., Hillenbrand, L.~A., Backman, D., et al.\ 2006, \pasp, 118, 1690 

Narayan, R., Cumming, A., \& Lin, D.~N.~C.\ 2005, ApJ, 620, 1002

Nelson, A.~F., \& Angel, J.~R.~P.\ 1998, ApJ, 500, 940 

Noyes, R.~W., Hartmann, L.~W., Baliunas, S.~L., Duncan, D.~K., \& Vaughan, A.~H.\ 1984, ApJ, 279, 763

Paulson, D.B., Cochran, W.D., \& Hatzes, A.P. 2004, AJ, 127, 3579 

Saar, S.~H., \& Donahue, R.~A.  \ 1997, ApJ, 485, 319

Santos, N.~C., Gomes da Silva, J., Lovis, C., \& Melo, C.\ 2010, A\&A, 511, A54 


White, R.~J., Gabor, J.~M., \& Hillenbrand, L.~A.\ 2007, AJ, 133, 2524 

Wright, J.~T., Marcy, G.~W., Howard, A.~W., et al.\ 2012, ApJ, 753, 160 

\end{references}
\end{document}